\documentclass[pre,twocolumn,showpacs,preprintnumbers,amsmath,amssymb,superscriptaddress,10pt]{revtex4}
\usepackage{graphicx}% Include figure files
\usepackage{dcolumn}% Align table columns on decimal point
\newcommand{\be}{\begin{equation}}
\newcommand{\ee}{\end{equation}}
\newcommand{\bd}{\begin{displaymath}}
\newcommand{\ed}{\end{displaymath}}
\newcommand{\BE}{\begin{eqnarray}}
\newcommand{\EE}{\end{eqnarray}}

\newcommand{\bz}{\ensuremath{\mathbf{z}}}

\newcommand{\bn}{\ensuremath{\mathbf{n}}}

\newcommand{\bEta}{{\mbox{\boldmath $\eta$}}}

\newcommand{\avg}[1]{\left\langle{#1}\right\rangle}

\begin{document}

%\preprint{APS/123-QED}

\title{Synchronization of coupled demographic oscillators}

\author{Tobias Galla}
\email{tobias.galla@manchester.ac.uk}
\affiliation{Theoretical Physics, School of Physics and Astronomy, The University of Manchester, Manchester M13 9PL, United Kingdom}

\date{\today}% It is always \today, today,
             % but any date may be explicitly specified

\begin{abstract}
Demographic oscillators are individual-based systems exhibiting temporal cycles sustained by the stochastic dynamics of the microscopic interacting particles. We here use the example of coupled predator-prey oscillators to show that synchronization to a common frequency can occur between two such systems, even if they oscillate at different frequencies in the absence of coupling. The power spectra of the separate and the coupled systems are computed within a van Kampen expansion in the inverse system size, and it is found that they exhibit two peaks at separate frequencies at low coupling, but that only one peak is present at large enough coupling strength. We further make predictions on the time behaviour of the phases of the two oscillators, and their phase difference, and confirm the frequency entrainment at sufficiently large coupling. Theoretical results are verified convincingly in numerical simulations.
 
\end{abstract}
\pacs{05.45.Xt; 02.50.Ey, 87.23.Cc}

                              %display desired
\maketitle

%%%%%%%%%%%%%%%%%%%%%%%%%%%%%%%%%%%%%%%%%%%%%%%%%%%%%%%%%%%%%%%%%%%%%%%%%%%%%%

Coupled systems of self-sustained or driven oscillators are abundant in physics, chemistry, biology and other adjacent sciences \cite{kurths}, and their collective behavior is of great interest. One of the most studied phenomena is that of synchronization, i.e. the adjustment of rhythms of oscillating units in the presence of weak interaction. Even if each oscillator in separation may have their own eigenfrequency, the ensemble of units starts oscillating at common frequency provided the coupling between them exceeds a threshold value. This phenomenon known as frequency entrainment or phase locking \cite{kurths}. 

Synchronization effects of this type have been observed in a variety of contexts, going back to Christiaan Huygens in 1673 \cite{huygens,kurths,chaos}, and ranging from coupled pendula and clocks, to Josephson junctions, neural activity, pacemaker cells in the sino-atrial node to predator-prey cycles (see \cite{kurths} and references therein).

A variety of different mathematical models of oscillating units, and of the interactions between them has been considered in the literature, including coupled non-linear oscillators described by deterministic differential equations as e.g. in the celebrated Kuramoto model {\cite{kuramoto}, time-delayed differential equations \cite{delay}, discrete systems \cite{wood}, chaotic oscillators \cite{chaos}, stochastic oscillatory systems driven by or subject to external noise \cite{freund,blasius}, and more recently individual-based spatial models in which diffusion plays the role of interaction \cite{efimov}.

In the present work we will focus on synchronization of oscillators driven by so-called demographic stochasticity \cite{nisbet}. These are individual-based models, in which the constituting reagents (the individuals) interact by a set of simple stochastic processes resulting in the creation and removal of individuals, or in conversion of an individual of one type into an individual of another type. Chemical reactions are an example of such systems, but the dynamics of agents in the context of game theory or the interactions in predator-prey systems \cite{alan} can be considered in the same framework. Depending on model parameters the mean-field dynamics of such systems may or may not allow for periodic solutions. Assuming a regime in which the deterministic {\em infinite} system does not exhibit oscillations, cyclic behavior can still be found in a population of a {\em finite} number of individuals operating at the same model parameters \cite{alan}. The random nature of the interaction between individuals is a source of stochasticity (so-called demographic stochasticity or intrinsic noise), and may under suitable circumstances lead to coherent sustained oscillatory behavior via a mechanism of resonant amplification. These effects have been observed in a number of different individual based models \cite{various}, and we will refer to such systems as `demographic oscillators' in the following.

Specifically, we will consider two coupled predator-prey systems, as studied in isolation in \cite{alan}. The two sub-systems will be labelled by $i=1,2$, each containing two types of individuals, predators and prey. We will label predators in system $i=1$ by $A_1$ and predators in system $i=2$ by $A_2$, and similarly we will use $B_1$ and $B_2$ for the prey. In addition both systems may contain vacancies $E_1$ and $E_2$. We will assume that each system contains $N$ individuals (including vacancies), i.e. that the number of $A_1$, $B_1$ and vacancies $E_1$ sum to $N$, and similarly for the second system. Following \cite{alan} the following processes define the dynamics of each of the sub-systems:
\BE
B_i+E_i&\stackrel{\beta_i}{\longrightarrow}&B_i+B_i\nonumber \\
A_i&\stackrel{\alpha_i}{\longrightarrow}&E_i\nonumber \\
B_i&\stackrel{\gamma_i}{\longrightarrow}&E_i\nonumber \\
A_i+B_i&\stackrel{\nu_{i}}{\longrightarrow}&A_i+A_i\nonumber \\
A_i+B_i&\stackrel{\mu_{i}}{\longrightarrow}&A_i+E_i.\label{eq:react1}
\EE
The $\{\beta_i, \alpha_i,\gamma_i,\nu_i,\mu_i\}$ are here rate constants ($i=1,2$). The first reaction describes a birth process, the second and third reactions are death processes, and the remaining two reactions in each system represent predator-prey interaction. Up to this point the two systems $i=1,2$ are entirely separate and it was shown in \cite{alan} that they exhibit demographic oscillations provided the reaction rates are chosen appropriately. In particular the power spectra of these oscillations was obtained analytically, and as shown in \cite{alan} they exhibit a peak at a characteristic frequency set by the reaction rates.

We will in the following focus on the case in which the reaction rates of the two oscillators are chosen such that in separation both systems oscillate at {\em distinct} frequencies. A coupling between the two systems is the introduced by allowing for the exchange of prey individuals according to 
\BE
B_1+E_2&\stackrel{\varepsilon}{\longrightarrow}& B_2+E_1,\nonumber \\
B_2+E_1&\stackrel{\varepsilon}{\longrightarrow}&B_1+E_2.
\EE
$\varepsilon$ hence defines the strength of the coupling between the two oscillators. It is worth pointing out that the number of individuals (including vacancies) in each system is constant in time under the reaction dynamics.

For later convenience, we will denote the state of the coupled system by the vector $\bn=(n_1,n_2,n_3,n_4)$ in the following. The $n_a,\, a=1,\dots,4$ are integers, and $n_1$ describes the number of predators in the first system, $n_2$ the number of prey in the first system, and $n_3$ and $n_4$ refer to predators and prey in system two. The dynamics is inherently stochastic and the time-evolution is governed by a master equation of the form
\be\label{eq:master}
\frac{dP(\bn,t)}{dt}=\sum_{\bn'} [T(\bn|\bn')P(\bn',t)-T(\bn'|\bn)P(\bn,t)],
\ee 
where $T(\bn|\bn')$ is the transition rate from state $\bn'$ to $\bn$. These rates are defined by the set of allowed reactions, and we will not report their detailed mathematical form.

Similar to \cite{alan}, the mean field limit results in deterministic equations for the mean values $x_1=n_1/N$, $y_1=n_2/N$, $x_2=n_3/N$ and $y_2=n_4/N$. For the coupled system these equations are of the form
\BE
\dot x_1&=&2\nu_{1}x_1y_1-\alpha_{1}x_1 \nonumber \\
\dot y_1&=&2\beta_1(1-x_1-y_1)y_1-\gamma_1y_1-2(\nu_1+\mu_1)x_1y_1\nonumber \\
&&+\varepsilon\left[y_2(1-x_1-y_1)-y_1(1-x_2-y_2)\right] \nonumber \\
\dot x_2&=&2\mu_{1}x_2y_2-\gamma_{2}x_2  \nonumber \\
\dot y_2&=&2\beta_2(1-x_2-y_2)y_2-\gamma_2y_2-2(\nu_2+\mu_2)x_2y_2\nonumber \\
&&+\varepsilon\left[y_1(1-x_2-y_2)-y_2(1-x_1-y_1)\right].\label{eq:determ}
\EE
As the only coupling between the two predator-prey system is through the exchange of prey individuals, the equations for the pairs $(x_1,y_1)$ and $(x_2,y_2)$ respectively decouple for $\varepsilon=0$. As discussed in \cite{alan} deterministic systems of this type approach a non-trivial fixed point asymptotically at $\varepsilon=0$ and do not allow any limit-cycle solutions.

The above mean-field equations are however valid only in the limit of infinite systems, $N\to\infty$. At finite sizes coherent amplified oscillations about the mean-field fixed points have been observed \cite{alan}, and their power spectra have been computed within a systematic expansion in the inverse system size. The spectra exhibit peaks at a non-zero frequency, determined by the rate constants and setting the characteristic period of the stochastic oscillations. We will focus on cases in which the two oscillators $i=1,2$ have characteristically different frequencies of their respective demographic oscillations. Specifically we will use $(\alpha_1,\beta_1,\gamma_1,\nu_1,\mu_1)=(0.1, 0.1, 0.0, 0.25,0.05)$ and $(\alpha_2,\beta_2,\gamma_2,\nu_2,\mu_2)=(0.1, 0.1, 0.1, 0.5,0.5)$. In separation the two systems show cyclic behavior as shown in the lower two panels of Fig. \ref{fig:timeseries}.
\begin{figure}[t]\vspace{1em}
\centerline{\includegraphics[width=0.5\textwidth]{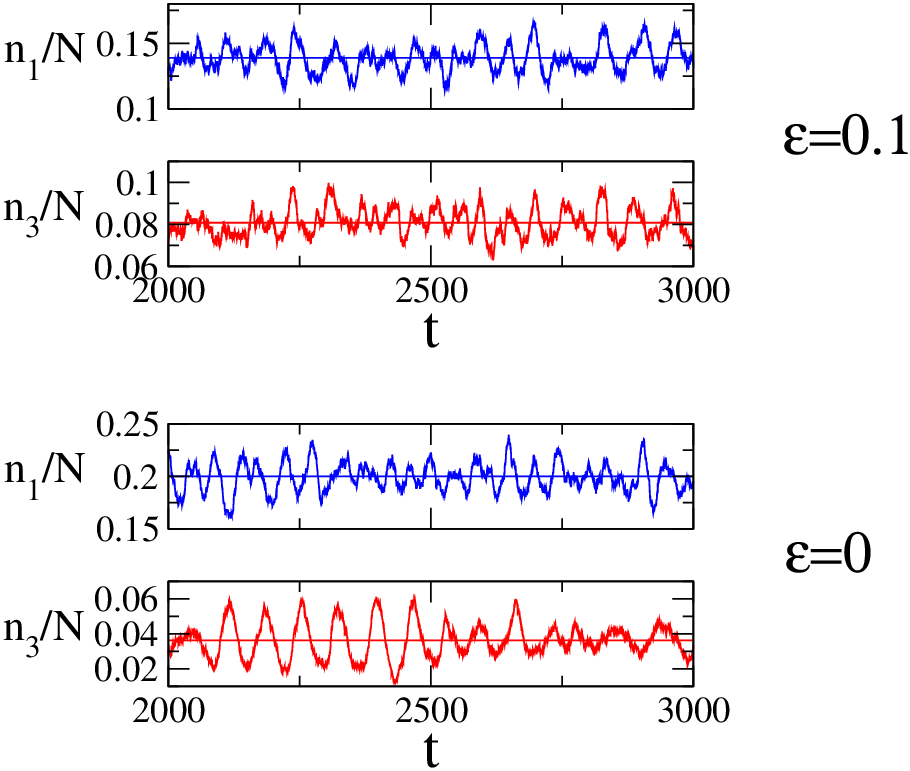}}\vspace{1em}
\caption{Predator densities in the two subsystems as a function of time. Upper two panels: the two oscillators are coupled and synchronize; lower two panels: uncoupled case, no synchronization. Horizontal lines in each panel represent the fixed-point solution of the deterministic dynamics Eqs. (\ref{eq:determ}), noisy curves show a single simulation run of the stochastic system with $N=5000$ individuals in each oscillator.}
\label{fig:timeseries}
\end{figure}
The periodic behavior of the system at finite sizes can be investigated analytically within a system-size expansion as proposed first by van Kampen \cite{kampen}. In essence the method consists in expanding about the mean-field solution, writing e.g. $n_1/N=x_1+z_1/\sqrt{N}$ and similarly for the remaining three components of the system. Changing from the variables $n_1,n_2,n_3,n_4$ to $z_1,z_2,z_3,z_4$ in the master equation one then performs a systematic expansion in powers of $N^{-1/2}$. We will not present the details of the mathematics here, as these have been reported for similar systems in the literature \cite{kampen,alan,various}, but will only report the final result. The leading order of this expansion reproduces the mean-field equations (\ref{eq:determ}), as expected. In next-to-leading order one obtains a Langevin equation 
\be
\dot\bz=J\bz+\bEta(t)
\ee
for the vector $\bz=(z_1,z_2,z_3,z_4)$ describing fluctuations about the mean-field result. $\bEta$ is here a $4$-component Gaussian noise vector, which comes out as white in time, but with correlations between its components, as we will specify below. Focusing on the asymptotic regime in which the mean-field theory attains a fixed point, $J$ is the $4\times 4$ Jacobian of the system (\ref{eq:determ}) at this fixed point \cite{alan}. The correlator of the noise variables can be worked out and one finds
\be
\avg{\eta_a(t)\eta_b(t')}=\delta(t-t')\sum_{r=1}^{12} a_r (\delta n_a)_r (\delta n_b)_r
\ee
for the case of two coupled predator-prey systems. $a,b\in\{1,\dots,4\}$ label the components of the noise, and the index $r$ runs through the twelve possible reactions in the system (five in each subsystem, and the two exchange reactions), $a_r$ stands for the rate at which reaction $r$ occurs in the mean-field system. The integer numbers $(\delta n_a)_r$ finally denote the number of particles of type $a$ that are created ($(\delta n_a)_r>0$), or removed from the system ($(\delta n_a)_r<0$) during one occurrence of reaction $r\in\{1,\dots,12\}$. If we give reaction $B_1+E_1\stackrel{\beta_1}{\rightarrow} B_1+B_1$ the label $r=1$ then one has $a_1=\beta_1 y_1 (1-x_1-y_1)$, and $(\delta n_1)_1=0,(\delta n_2)_1=1,  (\delta n_3)_1=(\delta n_4)_1=0$ for example. Analogous expressions can be formulated for the remaining reactions.
\begin{figure}[t]\vspace{1em}
\centerline{\includegraphics[width=0.4\textwidth]{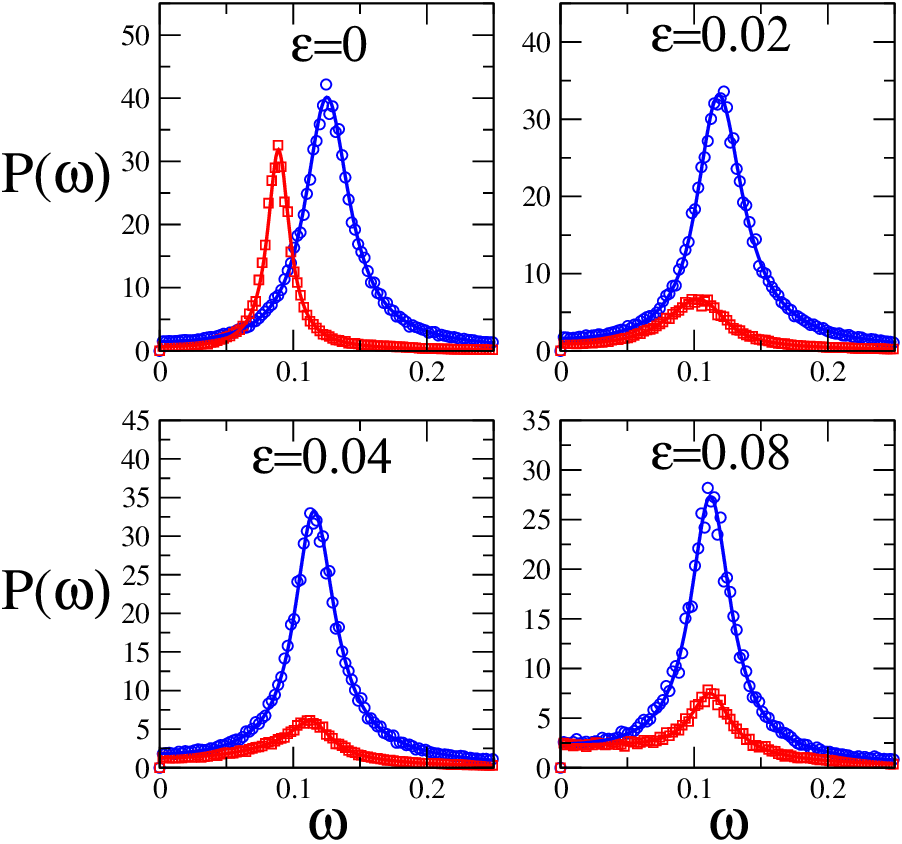}}\vspace{1em}
\caption{(color on-line) Power spectra $P_{11}(\omega)$ (blue circles) and $P_{33}(\omega)$ (red squares) of the predator densities in the two sub-systems at different magnitudes of the coupling strength $\varepsilon$. Solid lines are obtained from the theory, markers from simulations of the individual-based model ($N=10^4$, averages over at least $200$ samples are taken in each panel).}
\label{fig:spectra}
\end{figure}
The above Langevin equation for $\bz$ is linear, and can be solved in Fourier space. Details for similar systems are again found in the literature, so that we do not describe the details of the intermediate steps here. The outcome is the set of power spectra
\be
P_{ab}(\omega)=\avg{\widetilde z_a(-\omega)\widetilde z_b(\omega)},~~~a,b\in\{1,\dots,4\},
\ee
where $\widetilde z_a(\omega)$ is the Fourier transform of $z_a(t)$, and where $\avg{\cdots}$ stands for an average over realizations of the Langevin dynamics.
Results from this analytical computation are shown in Fig. \ref{fig:spectra}, where we plot the power spectra of the predator densities in either of the two sub-systems, $P_{11}(\omega)$ and $P_{33}(\omega)$. The solid lines are the predictions of the theory, markers are from simulations based on the Gillespie algorithm \cite{gillespie}. As seen in the figure the agreement between theory and numerical experiment is excellent. With our choice of parameters the dominating frequency components in the fluctuations of the number of predators are quite distinct in the absence of coupling between the systems. Two peaks at distinct frequencies are seen in the top left panel of Fig. \ref{fig:spectra}. As we gradually increase the coupling $\varepsilon$ between the two demographic oscillators, the positions of the two peaks approach each other, and finally the dominant frequencies of both systems essentially coincide at large enough $\varepsilon$ (see Fig. \ref{fig:spectra}), leading to frequency entrainment. This effect is also visible in the time series shown in Fig. \ref{fig:timeseries}. While the fluctuations in the number of predators have characteristically distinct periods in the absence of coupling between the two oscillators (lower panel), coherent oscillations at a unique common frequency are found for $\varepsilon=0.1$ (upper panel). Due to the stochasticity in the oscillations this coherence can never be expected to be perfect, but the qualitative effect is certainly visible in the figure.

To further illustrate the synchronization effect in the system we have monitored the position of the peaks in the power spectra (as obtained from the system-size expansion) as $\varepsilon$ is varied. Results are shown in the upper panel of Fig. \ref{fig:eigen}, and while the peaks are well separated in the regime of small coupling, their peak positions essentially coincide for values of the coupling strength greater than about $\varepsilon=0.07$. While this is not an analytically exact result, the deviation between the position of the peaks is minute in this regime, and for all practical purposes it is therefore fair to conclude that the two systems have indeed synchronized. 

Further insight can be gained by computing the eigenvalues of the Jacobian $J$ at the fixed-point of the deterministic system. The presence of sustained demographic oscillations in the stochastic system is here indicated by non-zero imaginary parts of these eigenvalues. In essence, non-vanishing imaginary parts cause an oscillatory approach of the deterministic system to the stable fixed point (real parts of the eigenvalues are negative). In the finite system the stochastic nature of the microscopic dynamics results in an ongoing random perturbation away from this fixed point. An instantaneous perturbation would decay in an oscillatory manner back to the fixed point, but as perturbations due to the demographic noise occur at all times, a coherent oscillatory pattern emerges. In a crude approximation one may estimate that the frequency of such oscillations is set by the imaginary parts of the eigenvalues of the Jacobian. Our system is $4$-dimensional, so we have four eigenvalues of the Jacobian of the coupled system, and three different scenarios are possible: (i) all four eigenvalues are real (no demographic oscillations), (ii) they form two distinct pairs of complex conjugates (resulting in two distinct characteristic frequencies), or (iii) two of them are real, and the remaining two are complex conjugates with non-zero imaginary part (resulting in only one characteristic frequency). As seen in the lower panel of Fig. \ref{fig:eigen}, (ii) is the case at low coupling strengths, up to about $\varepsilon=0.11$ for the specific reaction rates chosen throughout this paper. Above this critical coupling strength, only one pair of complex-conjugate eigenvalues is found, indicating that there is only one characteristic frequency in the system. In case of two pairs of complex-conjugates, care needs to be taken however in identifying the resulting frequencies with those of oscillations of the various species concentrations in the system. The precise eigenvector structure may here be important if one wanted to identify the relevant normal modes (no such attempt has been made here), and
\begin{figure}[t]\vspace{1em}
\centerline{\includegraphics[width=0.4\textwidth]{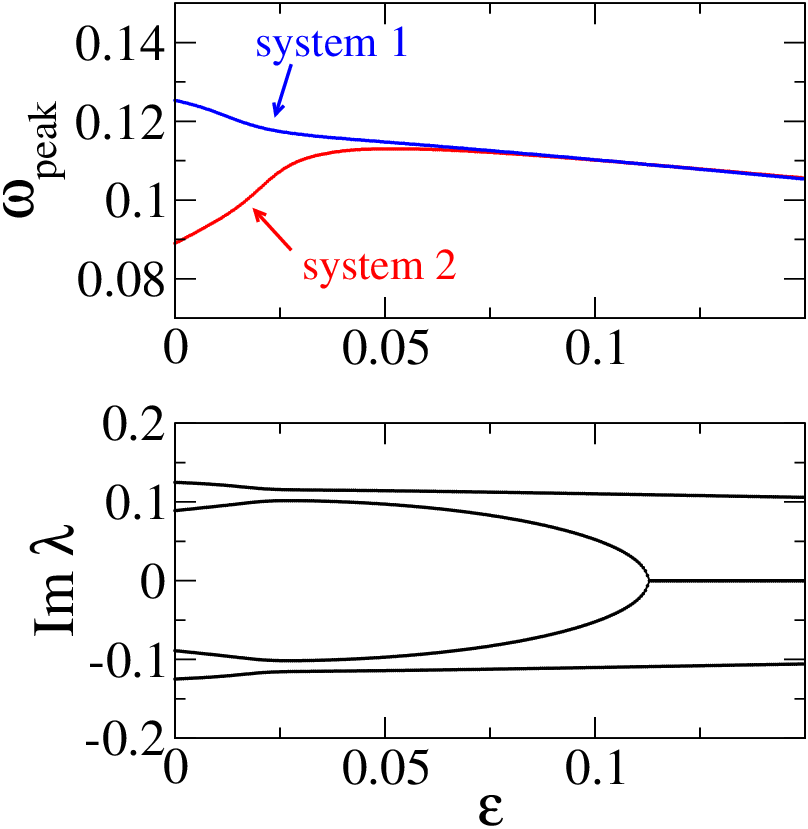}}\vspace{0.5em}
\caption{(color on-line) Position of the peaks of the power spectra of predator fluctuations (upper panel) as a function of the coupling strength $\varepsilon$. The lower panel shows the imaginary parts of the eigenvalues of the Jacobian $J$ (see text for details). }\vspace{0em}
\label{fig:eigen}
\end{figure}
we must stress that the peak of the different power spectra is generally
not located precisely at the frequencies set by the imaginary parts of
the eigenvalues, but that their real parts and the correlations
between the noise components $\eta_a$ may have an impact on their
position as well. 

To further study the phase synchronization of the two demographic
oscillators we have measured their relative phases numerically, and
report results in Fig. \ref{fig:phase}. The phase $\phi(t)$ of a
narrow frequency-band periodic signal is here defined so that it
increases by $2\pi$ within one oscillation cycle, and grows in
proportion to the fraction of the period that has elapsed
\cite{kurths}.
\begin{figure}[t!!!]\vspace{4em}
\centerline{\includegraphics[width=0.4\textwidth]{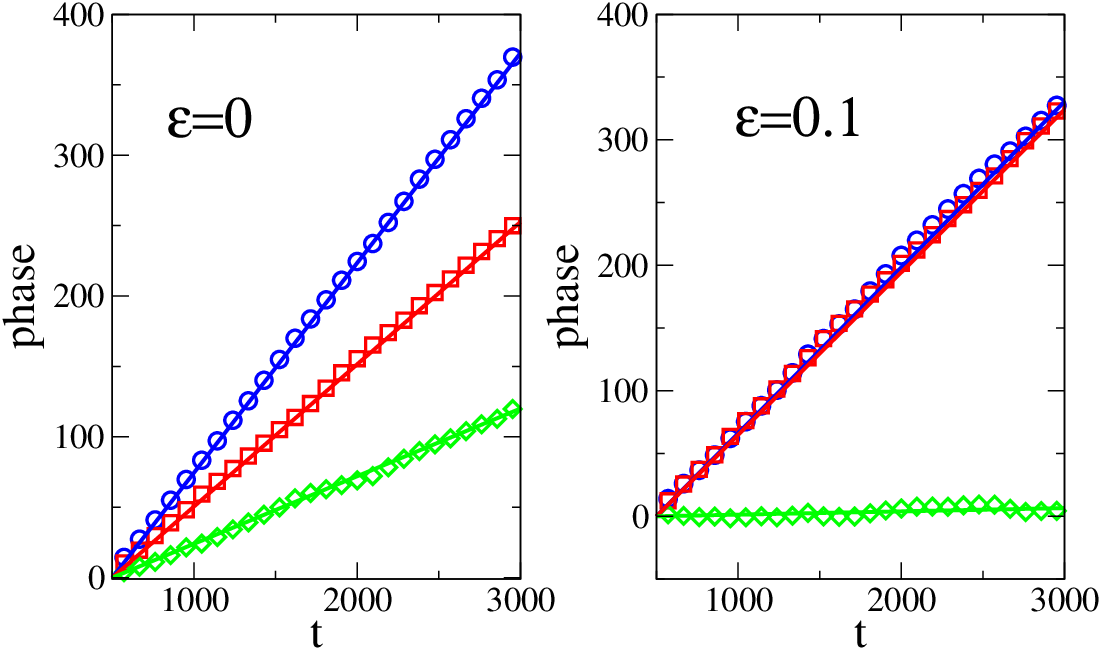}}
\caption{(color on-line) Instantaneous phases of the predator fluctuations in the two oscillators. Symbols are from simulations (circles are $\phi_1$, squares $\phi_3$, diamonds $\phi_1-\phi_3$; $N=10^4$, averaged over $20$ samples), solid lines have slopes $\omega=\int_0^{\Omega}d\omega' \omega' P_{aa}(\omega')/[\int_0^{\Omega}d\omega' P_{aa}(\omega')]$, where $P_{aa}(\omega)$ ($a\in\{1,3\}$) is the analytically computed spectrum. $\Omega\approx 1.2$ is a cutoff used to reduce the effects of phase slips \cite{hilbert}. Measurements of the phase start at $t=500$, to allowing for equilibration.}\vspace{1em}
\label{fig:phase}
\end{figure}
Extracting the instantaneous phase $\phi(t)$ from a
stochastic signal generated by Gillespie simulations is
non-trivial, we have here resorted to Hilbert transform
techniques as described in \cite{kurths}, note also \cite{hilbert}. As
seen in the left panel of Fig. \ref{fig:phase} the analysis of the phases of the two demographic systems confirms that they oscillate at distinctively different frequencies if uncoupled, and the difference in their phases increases linearly in
time. At sufficiently large coupling (right panel) the phases of both sub-systems grow at the same slope in time (the slopes of the curves in the right panel of Fig. \ref{fig:phase}) coincide within a margin of about $2$ per cent), and the phase difference hence remains finite in time, indicating phase locking and frequency entrainment.
\begin{figure}[t!!!]\vspace{4em}
\centerline{\includegraphics[width=0.3\textwidth]{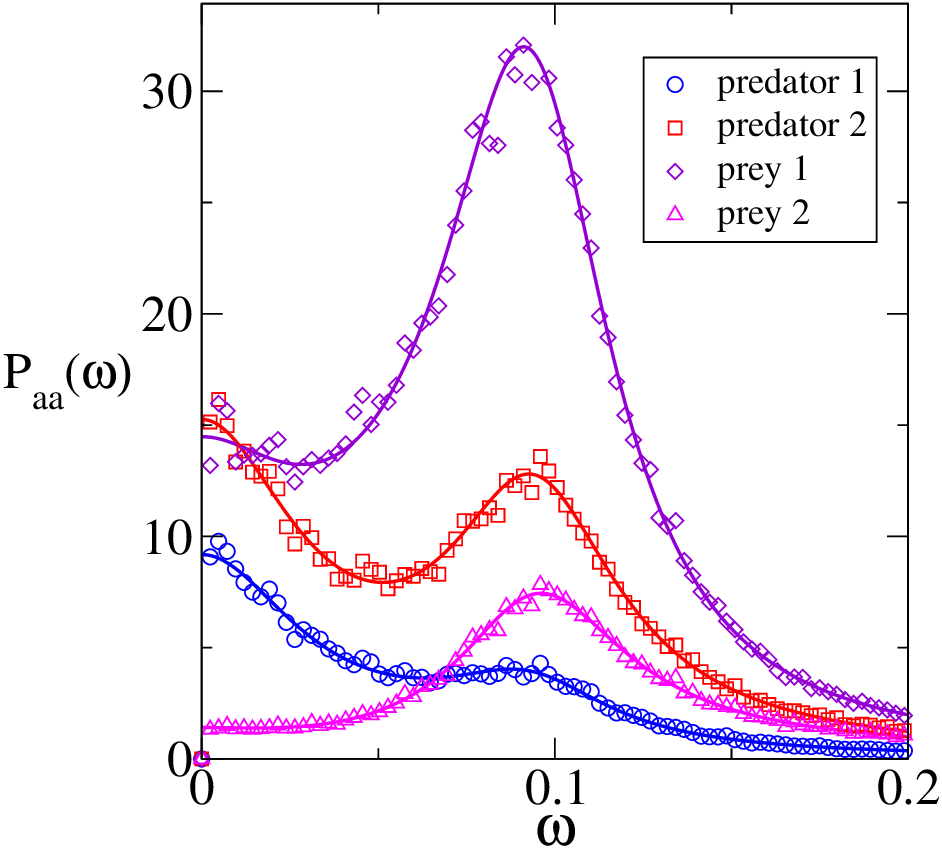}}
\caption{(color on-line) Power spectra of the fluctuations of the four species at $\varepsilon=0.3$. Solid lines are analytical results obtained within the van Kampen expansion, markers from simulations at $N=10^4$ (averaged over $500$ samples).}\vspace{1em}
\label{fig:eps}
\end{figure}

Before concluding, we would like to point out that a systematic investigation of the behaviour of the coupled predator-prey system at values of the reaction rates different from the choice made in the present work is still pending, and no claim is made here that synchronization will occur for {\em all} choices of the reaction rates in (\ref{eq:react1}). Indeed as shown in Fig. \ref{fig:eps} oscillation death \cite{strogatz} may occur at sufficiently large coupling strength, i.e. the (global) maximum of the power spectra of individual species may no longer be located at a non-zero frequency, indicating that demographic oscillations no longer exists for certain species in the system. For other choices of the reaction rates, this may in principle occur before synchronization is reached, so that such a system would not exhibit phase locking, but rather oscillation death would preempt synchronization of demographic oscillations under such circumstances.

In summary we have shown that the paradigm of synchronization at sufficiently large coupling extends to demographic oscillators, i.e. system in which the periodic behavior in each unit is sustained by coherent amplification of demographic noise. The model system we have focused on is that of two coupled predator-prey systems, and the synchronization effects we find indicate that the entrained dynamics of lynx abundances in different regions of Canada as reported in \cite{elton} may be explained within the picture of quasi-cycles driven by intrinsic noise (see also \cite{blasiusnature} and references therein for theoretical models of synchronization in spatially extended ecological systems). We expect that a similar mechanism will occur in other demographic oscillators, for example in biochemical reactions \cite{kiss} or in models of epidemics \cite{various}, and in systems which are composed of more than two self-sustained oscillators.

{\em Acknowledgments:} TG is an RCUK Fellow (RCUK reference EP/E500048/1), and would like to thank A. J. McKane and R. P. Boland for useful discussions on demographic oscillators.


\begin{thebibliography}{53}
\bibitem{hilbert} To extract the instantaneous phase we use a procedure similar to that described in \cite{kurths}. Prior to performing the Hilbert transform we have eliminated high frequency modes ($\omega>\Omega\approx1.2$) in order to smoothen the signal and to minmise effects of phase slips. The slopes reported in Fig. \ref{fig:phase} do depend on the choice of $\Omega$, but agreement between theory and simulations remains as $\Omega$ is varied, provided the cutoff is not too large.
\end{thebibliography}
\end{document}